\documentclass[]{wan}
\usepackage{wrapfig}
\usepackage{tabularx}
\usepackage{textcomp}
\usepackage{stfloats}
\usepackage{url}
\usepackage{verbatim}
\usepackage{graphicx}
\usepackage{titlesec}
\usepackage{tocloft}
\usepackage{adjustbox}
\usepackage{multirow}
\usepackage{pifont}
\usepackage{tikz}
\usepackage{comment}
\usepackage{amsmath,amssymb} % define this before the line numbering.
\usepackage{colortbl}  
\usepackage{color}
\usepackage{booktabs} 
\usepackage{hyperref}
\usepackage{graphicx}   
\usepackage{subcaption} 
\usepackage{booktabs}
\usepackage{amsmath} % 
\usepackage{multirow} % 
\usepackage{booktabs} % 
\usepackage{subcaption} %  
\usepackage{graphicx}   % 
\usepackage{wrapfig}    % 
\usepackage{caption}    % 
\usepackage{attachfile2}

\NewDocumentCommand{\audiobutton}{O{\faMusic} O{Click to play} O{blue!70} m m}{%
   \textattachfile[color=1 1 1, mimetype=audio/mp3]{#4}{%
     \colorbox{#3}{\color{white}#1 ~~#2 \ \raisebox{-0.25em}{\includegraphics[height=1.2em]{#5}}}%
   }%
}
\RequirePackage{xspace}
\makeatletter
\DeclareRobustCommand\onedot{\futurelet\@let@token\@onedot}
\def\@onedot{\ifx\@let@token.\else.\null\fi\xspace}

\def\eg{\emph{e.g}\onedot} 

\def\ie{\emph{i.e}\onedot}

\def\etc{\emph{etc}\onedot}

\makeatother

\usepackage{makecell}
\definecolor{Gray}{gray}{0.95}

\definecolor{adptorange}{RGB}{248, 205, 172}
\definecolor{cmpblue}{RGB}{189, 215, 238}
\definecolor{cmpblue}{RGB}{189, 215, 238}

\definecolor{our_red}{RGB}{232,157,160}
\definecolor{our_blue}{RGB}{136,206,230}
\definecolor{our_orange}{RGB}{246,200,168}
\definecolor{our_green}{RGB}{178,211,164}

\definecolor{attn_code0}{RGB}{247,215,200}
\definecolor{attn_code1}{RGB}{238,169,139}
\definecolor{mlp_code0}{RGB}{204,201,221}
\definecolor{mlp_code1}{RGB}{102,95,153}

\definecolor{token_blue}{RGB}{84, 120, 140}
\definecolor{myMagenta}{rgb}{0.9,0,0.4}

\usepackage{pifont}       % \ding{xx}
\usepackage{bbding}       % \Checkmark,\XSolid,... (需要和pifont宏包共同使用)
\usepackage{fontawesome}
\usepackage{xspace}

\usepackage{float}

\newcommand{\tablestyle}[2]{\setlength{\tabcolsep}{#1}\renewcommand{\arraystretch}{#2}\centering\footnotesize}

\newcolumntype{x}[1]{>{\centering\arraybackslash}p{#1pt}}
\newcolumntype{y}[1]{>{\raggedright\arraybackslash}p{#1pt}}
\newcolumntype{z}[1]{>{\raggedleft\arraybackslash}p{#1pt}}

\usepackage{colortbl}
\usepackage{xcolor}
\usepackage{wrapfig}

\renewcommand{\paragraph}[1]{\vspace{1.25mm}\noindent\textbf{#1}}

\usepackage{algorithm}
\usepackage{listings}

\definecolor{codeblue}{rgb}{0.25, 0.5, 0.5}
\definecolor{codekw}{rgb}{0.35, 0.35, 0.75}
\lstdefinestyle{Pytorch}{
    language = Python,
    backgroundcolor = \color{white},
    basicstyle = \fontsize{9pt}{8pt}\selectfont\ttfamily\bfseries,
    columns = fullflexible,
    aboveskip=1pt,
    belowskip=1pt,
    breaklines = true,
    captionpos = b,
    commentstyle = \color{codeblue},
    keywordstyle = \color{codekw},
}

\definecolor{green}{HTML}{009000}
\definecolor{red}{HTML}{ea4335}

\newcommand{\wan}{{Wan}\xspace}
\title{WanSong v1.0 Technical Report}

\author{Wan Team, Alibaba Group}

\newauthor{Binghui Chen\footnote{Project leader, email is chenbinghui@bupt.cn}, Pandeng Li, Yu Liu, Jingren Zhou}

\abstract{
Music generation foundation models have recently attracted significant industry attention. However, achieving efficient generation and high-fidelity long-form audio while supporting controllability remains challenging. To address these needs, we present \textbf{WanSong}, a simple yet powerful approach for long-form, commercial-grade song generation. Unlike autoregressive (AR) and cascaded multi-stage pipelines (\eg, AR followed by diffusion), \textbf{WanSong} is a pure diffusion-based model that directly generates high-fidelity, multilingual songs up to 5 minutes and outputs dual stems (vocals and background music) in a single run. In addition, our diffusion framework enables faster inference through step-distillation, and offers an efficient pathway for fine-tuning and customization to support downstream editing tasks.
}

\date{\today} 
\modelcard{WanSong v1.0}
\begin{document}
\makeatletter
\let\CT@arc@\relax
\let\CT@drsc@\relax
\makeatother
\thispagestyle{firstheader}
\maketitle
\pagestyle{empty}

\section{Introduction}
Music/song foundation models are undergoing a paradigm shift. Commercial-grade systems, \eg \cite{suno_discover} and \cite{mureka}, have demonstrated impressive music fidelity and fluency. However, most existing methods remain fundamentally based on autoregressive (AR) modeling. While AR models offer a principled way to generate sequences, they can lead to challenges in generation efficiency and in maintaining consistent perceptual quality for long-form audio.

Some recent works incorporate diffusion approaches on top of AR pipelines, for instance by using diffusion as an additional refinement module to improve audio realism. Building on these ideas, we further investigate how to design music foundation models in a more end-to-end manner. In particular, we aim to directly model audio generation without relying on a multi-stage AR-dominant pipeline, enabling a unified framework for controllable long-form song generation.

Motivated by this goal, we abandon conventional AR-dominant approaches and instead treat audio as a sequence of continuous tokens. We then develop an end-to-end, single-stage music foundation model using a pure diffusion scheme.

In this work, we present \textbf{WanSong v1.0}, an end-to-end single-stage diffusion-based music foundation model that treats audio directly as continuous tokens. Specifically, we employ a refined hybrid-MMDit architecture as the model backbone conditioned on an LLM text captioner. All text and audio tokens are sequentially concatenated into one unified sequence, feeding into the hybrid-MMDit model. However, we find that jointly encoding both vocals and BGM\footnote{or called accompaniment} within a single token causes the diffusion model to over-focus on the vocal components, while the more complex BGM part is suppressed. In addition, Classifier-Free Guidance (CFG) cannot achieve a good balance between vocals and BGM: as CFG increases, the vocals become more accurate but the BGM weakens; conversely, when CFG is smaller, the BGM improves but the vocals deteriorate. To address these issues, we propose a novel dual-stem token scheme that models vocals and BGM separately, fundamentally eliminating the interference between them. In addition, we also employ reinforcement learning (RL) to align the model with human preferences.

Our contributions are summarized as follows:
\begin{itemize}
    \item We present WanSong v1.0, a commercial-grade text-to-music model that supports multilingual, high-fidelity, and long-duration song generation.
    \item Pure diffusion-based framework: audios are treated as continuous tokens and then fed into an end-to-end hybrid-MMDit backbone for learning the relations between text captions and songs.
    \item Dual-stem modeling: We independently learn the vocal and BGM components, thereby preventing cross-interference between the two. As a result, WanSong can directly output separated stems for vocals and BGM, which makes subsequent post-production editing more convenient.
\end{itemize}
\section{Data}
\begin{figure}
  \centering
  \includegraphics[width=0.8\linewidth]{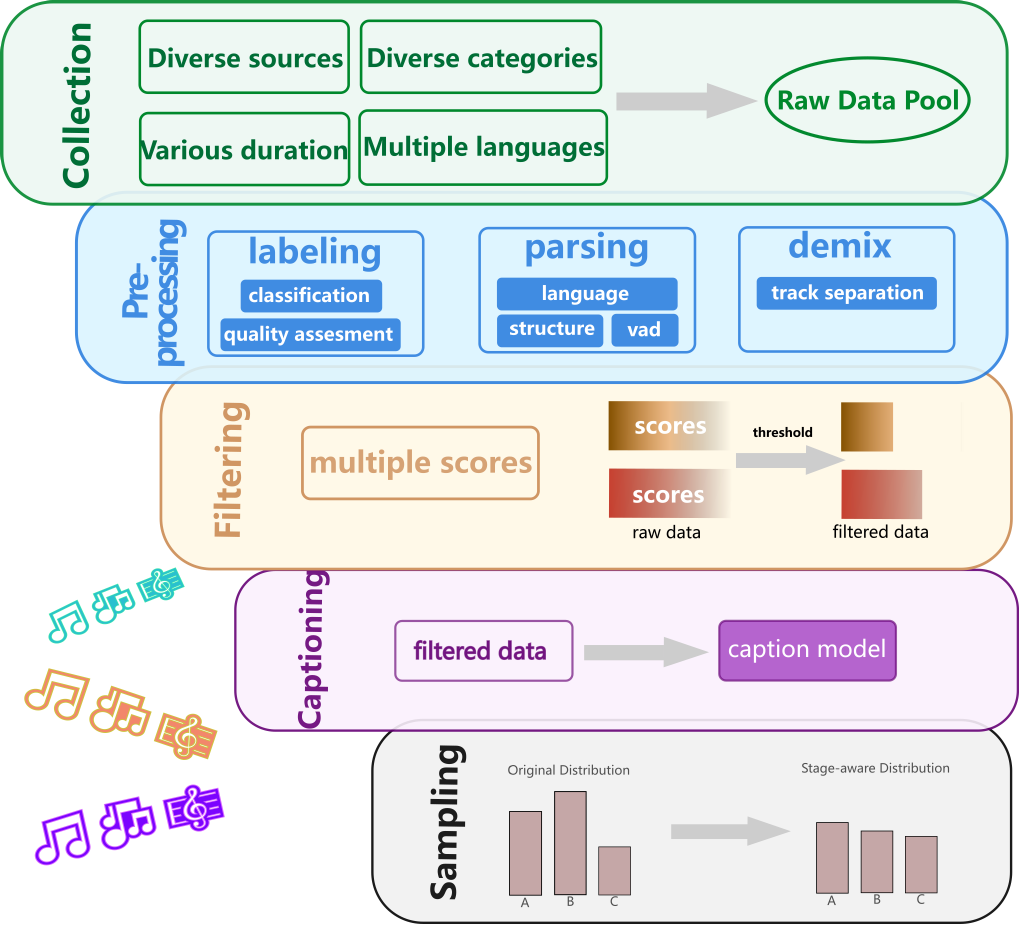}
  \caption{Data pipeline overview.}\label{fig_data}
\end{figure}

As a key foundation for large-scale music generative models, the quality and diversity of training data greatly determine overall model performance. To build a high-quality, well-balanced, and diverse dataset, we develop a rigorous data curation pipeline with five stages: collection, pre-processing, filtering, captioning, and sampling, as shown in Figure \ref{fig_data}.

\textbf{Data collection:} We collect a large and diverse corpus of audio data covering multiple languages, diverse musical styles, both vocal songs and instrumental music, and a wide range of song durations.

\textbf{Data pre-processing:} As the raw collected audio dataset is noisy, we implement a multi-stage pre-processing pipeline. Specifically, each audio file undergoes audio classification, quality assessment, structural parsing, VAD (voice activity detection), language classification, and track separation. This process yields a candidate pool of musics with diverse annotations.

\textbf{Data filtering:} By jointly considering the audio category and audio quality, we construct a high-quality, balanced sub-dataset for model training. In addition, the multi-dimensional labeling taxonomy we develop allows us to schedule training data in a flexible and adaptive manner across different training phases.

\textbf{Captioning:} Captioning is one of the most important components, and the effectiveness of captioning directly drives the model’s generation performance. In order to enhance the captioning ability, We have meticulously annotated musical style, vocals, instruments, mode and key, language, reverb, emotions, lyrics, and more. 

\textbf{Data sampling:} Since the collected data are imbalanced across different categories (\eg, music styles, vocal gender, audio quality and so on), we adopt a fine-grained balancing strategy at each stage of training.

The resulted training set has more than 6 million hours of multilingual song data.
\section{Method}
\subsection{Variational Autoencoder}

To bridge raw waveforms and the downstream diffusion transformer, we train a continuous 1-D Variational Autoencoder (VAE) that compresses a stereo audio signal into a compact latent stream. Compared to Stable Audio 2~\cite{stableaudio}, we halve the compression ratio to retain substantially more time-frequency detail, which is crucial for reproducing sharp transients and high-fidelity source separation.

\noindent\textbf{Overall Configuration.}
The VAE maps a stereo waveform $x \in \mathbb{R}^{2 \times T}$ at $44.1$\,kHz to a latent tensor $z \in \mathbb{R}^{C_z \times T'}$, where $C_z = 64$ and $T' = T / 1024$. This corresponds to a downsampling factor of $1024$ and a latent frame rate of $\approx 43.1$\,Hz. 

\noindent\textbf{Training Objective.}
The generator $G$ (encoder-decoder) and multi-scale discriminators $D$ are trained adversarially:
\begin{equation}
\begin{cases}
\mathcal{L}_{G} = \lambda_{\text{mag}}\mathcal{L}_{\text{mag}}  + \lambda_{\text{fmap}}\mathcal{L}_{\text{fmap}} + \lambda_{\text{adv}}\mathcal{L}_{\text{Adv}}(G; D) + \lambda_{\text{KL}}\mathcal{L}_{\text{KL}}, \\
\mathcal{L}_{D} = \mathcal{L}_{\text{Adv}}(D; G),
\end{cases}
\end{equation}
where:
\begin{itemize}
    \item $\mathcal{L}_{\text{mag}}$: Multi-resolution STFT magnitude loss computed on Mid-Side (M/S) and Left-Right (L/R) channels.
    \item $\mathcal{L}_{\text{fmap}}$: $\ell_1$ discriminator feature-matching loss.
    \item $\mathcal{L}_{\text{Adv}}$: Hinge adversarial loss.
    \item $\mathcal{L}_{\text{KL}}$: KL divergence with the standard normal prior.
\end{itemize}

\noindent\textbf{Implementation Details.}
We train the model on a curated corpus of $\sim 2.6 \times 10^{8}$ audio clips. We filter the dataset using quality classifiers to remove degraded, low-SNR, or low-bitrate samples, and apply silence detection to discard inactive segments. The retained corpus is explicitly balanced with a domain ratio of $\text{Music} : \text{Speech} : \text{Sound} = 50\% : 40\% : 10\%$. All clips are resampled to $44.1$\,kHz stereo, and we employ random phase inversion  for data augmentation.

We optimize the model using AdamW with a learning rate of $2\times 10^{-4}$ for $G$ and $3\times 10^{-4}$ for $D$. Training is conducted end-to-end on $32$ NVIDIA A100 GPUs for 1.5M steps using $1.5$\,s stereo crops. Notably, during the latter stage of training, we apply random noise augmentation to the latent representation to robustify the decoder against out-of-distribution artifacts during inference.

\subsection{Architecture}
All tokens, \ie, text tokens encoded by decoder-only LLM and dual-stem audio continuous tokens encoded VAE, are concatenated together as a sequence and then fed into our transformer-based diffusion backbone. It is worth noting that, in order to improve token throughput and token utilization within each batch, we pack tokens from different samples together by concatenating their sequences and feed the combined stream into the network.

\textbf{Hybrid transformer:} MMDit proposed by \cite{esser2024scaling} has shown promising results in image and video foundation model settings. We follow this basic idea to process tokens from different modalities. However, our experiments show that we do not need to learn separate parameters for different modalities in every layer. Inspired by the ideas in Flux (\cite{labs2025flux1kontextflowmatching}), we ultimately adopt a hybrid-transformer architecture. Moreover, following our previous settings in Wan2.1 (\cite{wan2025}), we employ the fully-shared AdaLN in each block to effectively reduce parameters while preserving performance.

\textbf{Dual-stem modeling}: In diffusion models, CFG typically serves to adjust the output distribution. However, in our early experiments, we found that a larger CFG improves phoneme accuracy, but it also suppresses the BGM by vocals. Conversely, a smaller CFG better restores the BGM, but the vocal accuracy degrades.
To address this issue, we propose a dual-stem output with joint-learning-in-layer strategy. It means that the vocal and BGM tokens are produced independently at the model’s output. While, during the learning of each block, we treat them as different channels of the same token, such that the model maintains independent output streams while simultaneously learning their intrinsic dependencies.

\begin{figure}
  \centering
  \includegraphics[width=0.95\linewidth]{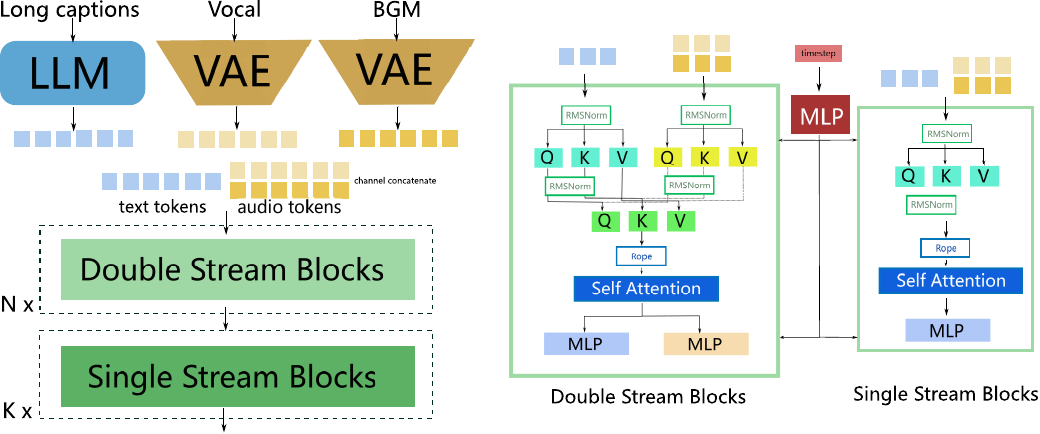}
  \caption{Our hybrid-transformer backbone.}\label{fig_arch}
\end{figure}

The overall architecture is shown in Figure \ref{fig_arch}. $\frac{N}{N+K}=0.3$ and the total model size is nearly 25B.
\subsection{Training}
To gradually learn the fine details of long-duration songs, the pre-training phase consists of 3 stages, including 90s-duration training, 300s-duration training and supervised-finetuning(SFT) stages. The flow matching framework (\cite{lipman2022flow}, \cite{esser2024scaling}) is used to model a denoising diffusion process within the audio domain. During training, given an audio latent $X_{1}$, a random noise $X_{0}\sim \mathcal{N}(0,I)$, and a timestep  $t\in[0,1]$ sampled from a logit-normal distribution, an intermediate latent $X_{t}$ is obtained as the training input. Following Rectified Flows (RFs) (\cite{esser2024scaling}), $X_{t}$ is defined as a linear interpolation between $X_{0}$ and $X_{1}$:
\begin{equation}
    X_{t}=tX_{1}+(1-t)X_{0}
\end{equation}
The ground truth velocity $V_{t}$ is:
\begin{equation}
    V_{t} = \frac{dX_{t}}{dt} =X_{1} - X_{0}
\end{equation}
Then the model can be trained to predict this velocity, the loss function can be formulated as the mean squared error (MSE) between the model output and $V_{t}$. In this work, our loss function is as:
\begin{equation}
    L =\alpha_{vocal}\mathbb{E}_{t}[||\theta(X_{vocal,t},C_{txt},t) - V_{vocal,t}||^{2}] + \alpha_{bgm}\mathbb{E}_{t}[||\theta(X_{bgm,t},C_{txt},t) - V_{bgm,t}||^{2}],
\end{equation}
where $\alpha_{vocal}$ and $\alpha_{bgm}$ are coefficients for balancing the learning strength between vocal and bgm. $C_{txt}$ represents the text condition. 
\subsection{Reinforcement Learning}
To further enhance the quality of generated songs/musics, we employ Reinforcement Learning from Human Feedback (RLHF) to align model with human preferences. Specifically, we focus on three critical dimensions that significantly influence the overall listening experience: \textit{\textbf{musicality}}, \textit{\textbf{lyric accuracy}} and \textit{\textbf{prompt alignment}}. For each dimension, we
collect dimension-specific prompts and infer out the corresponding audios, then human preference annotations over these audios are collected. We train reward models to provide fine-grained supervision signals. These reward models are subsequently used to guide the music generation model via reinforcement learning. Each reward model can be trained by:
\begin{equation}
    L = -\mathbb{E}_{a^{+},a^{-}}[\log\tau(R(a^{+})-R(a^{-}))]
\end{equation}
where $a^{+},a^{-}$ represent the positive and negative sample annotated by human, respectively. With these paired samples, the reward model can learn what is good and what is bad in each dimension. After training these reward models, we employ both DPO(\cite{wallace2024diffusion}) and ReFL(\cite{xu2023imagereward}). \footnote{The loss function details of DPO and ReFL are not the core of this paper, so please refer to the original paper for more details.}

Since the DPO scheme directly optimizes the relationships between positive and negative samples, it provides supervision for every timestep $t\in[0,1]$. This makes it more beneficial for optimizing the global structure of the audio. But a weakness of DPO is that it can’t precisely push both positive and negative samples toward more favorable directions. This can be seen directly from the loss function itself: the pairwise formulation inherently introduces a relatively large amount of noise. In contrast, ReFL-like methods might work well since it directly optimize the model by precise reward evaluation and information backward. Unfortunately, due to limitations in its formulation, ReFL only works during the low-noise stage. Therefore, it is inclined to optimize fine-grained details, and it is also more susceptible to being “hacked.” To achieve complementary effects, we first perform DPO training and then follow with ReFL training.
\section{Evaluation}
\subsection{VAE}
 
We evaluate reconstruction fidelity on two held-out benchmarks that jointly stress the two dominant modalities the VAE has to serve: a music benchmark of $200$ randomly held-out clips from the internal \wan-song corpus, and the standard SeedTTS speech benchmark~\cite{seedtts}, from which we randomly sampled $2000$ clips. 
For every clip we run encode $\to$ decode and report \textbf{(i)} STFT distance ($\ell_1$ on log-magnitude spectra, lower is better), \textbf{(ii)} mel-spectrogram distance (lower is better), and \textbf{(iii)} scale-invariant SDR (dB, higher is better). 

We compare against two baselines: the official Stable Audio 2 VAE (compression $2048$), and an ablation of our own model at compression $2048$ (``Ours 2048'') that shares our training data and network architecture — so the gap between ``Ours 2048'' and Stable Audio 2 isolates the effect of everything \emph{except} the compression-rate change, while the gap between ``Ours 1024'' and ``Ours 2048'' isolates the effect of the compression rate itself. Results are summarized in Table~\ref{table:vae_music} and Table~\ref{table:vae_speech}.

\begin{table}[!ht]
    \centering
    \caption{VAE reconstruction quality on the wan-song music benchmark (200 clips).}
    \tablestyle{6pt}{1.1} % Reduced padding to make the table smaller
    \resizebox{0.6\linewidth}{!}{ % Reduced size from 0.95 to 0.8
    \begin{tabular}{l|c|c|c}
    \hline
        Method & STFT $\downarrow$ & MEL $\downarrow$ & SI-SDR (dB) $\uparrow$ \\ \hline
        Stable Audio 2 & 1.430 & 0.856 & 4.386 \\ 
        Ours (2048) & 1.163 & 0.772 & 4.661 \\ \hline
        \textbf{Ours (1024)} & \textbf{1.029} & \textbf{0.629} & \textbf{7.246} \\ \hline
    \end{tabular}}
    \label{table:vae_music}
\end{table}

\begin{table}[!ht]
    \centering
    \caption{VAE reconstruction quality on the SeedTTS speech benchmark (2000 random samples).}
    \tablestyle{6pt}{1.1} % Reduced padding to make the table smaller
    \resizebox{0.6\linewidth}{!}{ % Reduced size from 0.95 to 0.8
    \begin{tabular}{l|c|c|c}
    \hline
        Method & STFT $\downarrow$ & MEL $\downarrow$ & SI-SDR (dB) $\uparrow$ \\ \hline
        Stable Audio 2 & 0.999 & 0.754 & 8.653 \\ 
        Ours (2048) & 0.783 & 0.593 & 10.371 \\ \hline
        \textbf{Ours (1024)} & \textbf{0.698} & \textbf{0.458} & \textbf{12.922} \\ \hline
    \end{tabular}}
    \label{table:vae_speech}
\end{table}

We present two key observations from these results:

\textbf{First, our architectural optimizations and domain balancing yield superior reconstruction over Stable Audio 2 at matched compression ratios.} Specifically, at a compression factor of $2048$, our model outperforms Stable Audio 2 across all metrics on both datasets (e.g., $-19\%$ STFT and $+0.28$\,dB SI-SDR on music; $-22\%$ STFT and $+1.72$\,dB SI-SDR on speech). We attribute these gains to our domain-balanced 260M corpus and wider encoder/decoder capacities. This validates our 50/40/10 domain rebalancing decision: a purely music-centric VAE systematically struggles with transient fricative and plosive speech details, whereas our joint music/speech/sound training resolves this weakness without compromising music-specific fidelity.

\textbf{Second, halving the compression ratio from 2048 to 1024 provides the most substantial boost in overall fidelity.} This change is most prominent in SI-SDR performance, triggering a $+2.86$\,dB improvement on the wan-song benchmark and a $+4.27$\,dB jump on SeedTTS. This demonstrates that a $2048\times$ latent grid is inherently too coarse to capture sharp, fine-grained transient details regardless of how heavily optimized the underlying neural backbone is, confirming the necessity of our $1024\times$ design choice.

\subsection{WanSong Bench}
\textbf{Objective quantitative results:} In order to evaluate the effectiveness of our WanSong, we collect a testing benchmark, including 200-samples covering four languages (Chinese, English, Japanese, Korean) and over ten Level-1 musical genres (\eg, Pop, Funk, EDM, Rock, Country, Jazz, \etc). All samples' durations are around 4 minutes. We evaluate the performances by Pronunciation Error Rate (PER), Songeval(\cite{yao2025songeval}), SongBench(\cite{wu2026songbench}) and Muq text alignment(\cite{zhu2025muq}). 

Moreover, since evalscore is trained on near 2,000 songs and SongBench is trained on near 11,000 songs, their generalization ability is difficult to guarantee. As shown in Table \ref{table:music}, the score differences are very small. To address this, we use our own \textbf{Musicality} evaluation model to perform the assessment. Specifically, this model employs a comprehensive scoring scheme that combines multiple aspects—melody, structure, and overall listening experience (“harmony,” “texture,” “arrangement,” “emotion,” “vocals,” and “reverb”). These components are weighted and aggregated with ratios of 0.4, 0.25, and 0.35, respectively, to produce the final single score. The score range is $[0-10]$. Our model are trained on 80k songs annotated by human. Finally, the overall performances comparisons are shown in Table \ref{table:music}.
\begin{table}[!ht]
\caption{Objective quantitative evaluations. The comparisons are conducted over WanSong, SunoV5.5(\cite{suno_discover}), SunoV5(\cite{suno_discover}), MurekaV7.6(\cite{mureka}) and Levo(\cite{lei2026levo})}
\resizebox{0.9\linewidth}{!}{
\begin{tabular}{lccccc}
\hline
Metrics               & \multicolumn{1}{c}{WanSong(ours)} & \multicolumn{1}{c}{SunoV5.5} & \multicolumn{1}{c}{SunoV5} & \multicolumn{1}{c}{MurekaV7.6} & \multicolumn{1}{c}{Levo} \\
\hline
\multicolumn{6}{c}{\emph{SongBench ($\uparrow$)}}                                                                                                                      \\
\hline
Melody                & \textbf{7.02}       &6.76              & 6.98                       & 6.86                           & 4.60                     \\
Arrangement           & \textbf{7.25}       &7.07              & 7.15                       & 7.09                           & 4.55                     \\
Musicality            & \textbf{6.17}       &5.99              & \textbf{6.17}              & 6.07                           & 3.96                     \\
Vocal                 & \textbf{7.38}       &7.13              & 7.25                       & 7.14                           & 5.2                      \\
Instrumental          & \textbf{7.24}       &6.99              & 7.12                       & 7.01                           & 5.05                     \\
Mixing                & \textbf{7.14}       &6.95              & 6.97                       & 6.82                           & 4.64                     \\
Structure             & \textbf{7.05}       &6.86              & 6.88                       & 6.77                           & 4.41                     \\
\hline
\multicolumn{6}{c}{\emph{SongEval ($\uparrow$)}}                                                                                                                       \\
\hline
Coherence             & \textbf{4.47}        & 4.43             & 4.44                       & 4.38                           & 3.42                     \\
Musicality            & 4.55                 & 4.28            & \textbf{4.56}              & 4.51                           & 3.59                     \\
Memorability          & \textbf{4.57}        & 4.41             & 4.53                       & 4.47                           & 3.47                     \\
Clarity               & \textbf{4.46}        & 4.33             & 4.41                       & 4.35                           & 3.45                     \\
Naturalness           & \textbf{4.4}         & 4.21             & 4.34                       & 4.32                           & 3.28                     \\
\hline
\multicolumn{6}{c}{\emph{AudioBox-Aesthetic ($\uparrow$)}}                                                                                                             \\
\hline
Content Enjoyment     & \textbf{7.59}      &7.49               & 7.49                       & 7.36                           & 7.30                     \\
Content Usefulness    & 7.85               &\textbf{7.91}      & 7.80                       & 7.57                           & 7.52                     \\
Production Complexity & \textbf{6.59}      &6.38               & 6.53                       & 6.50                           & 5.61                     \\
Production Quality    & \textbf{8.27}      &8.23               & 8.21                       & 8.03                           & 8.01                     \\
\hline
\multicolumn{6}{c}{\emph{Text alignment ($\uparrow$)}}                                                                                                                 \\
\hline
Muq                   & \textbf{0.44}      & \textbf{0.44}               & \textbf{0.44}              & 0.43                           & 0.34                     \\
\multicolumn{6}{c}{\emph{Musicality (ours) ($\uparrow$)}}                                                                                                              \\
\hline
Musicality            & \textbf{5.49}        & 4.31             & 4.18                       & 3.83                           & 1.69                     \\
\hline
\multicolumn{6}{c}{\emph{Lyric Following ($\downarrow$)}}                                                                                                                \\
\hline
PER(\%)               & \textbf{7.43}         & 9.86            & 22.80                      & 12.7                           & 27.11                   \\
\hline
\end{tabular}
}\label{table:music}
\end{table}

\textbf{Ablation study on compression ratio of token:} For convenience, we only conducted comparative experiments in the PT-90s stage. We trained our model based on these three different setups: (1) a VAE based on a 2048 compression rate; (2) a VAE based on a 1024 compression rate, with an additional patch size of 2 (\ie, concatenating adjacent tokens into a single token); and (3) a VAE based on a 1024 compression rate. We tested the above three models on our 90s-bench. The final results are shown in the table \ref{table_vae_comp}.

One can observe from Table~\ref{table_vae_comp} that increasing the token compression ratio generally leads to worse perceptual quality. 
When using a VAE with compression ratio 2048 (patch size = 1), the model achieves a PER of 19.2\% and a Quality score \footnote{Audio quality evaluation was conducted by our expert assessors, as both EvalScore~\cite{yao2025songeval} and AudioBox~\cite{tjandra2025aes} cannot precisely measure the song audio quality.} of 2.1. 
For the same effective compression ratio (Actual compression ratio = 2048), adopting a lower base compression ratio (1024) together with a larger patch size (patch size = 2, i.e., concatenating neighboring tokens) results in a slightly higher PER of 20.6\% and a similar Quality score of 2.4, suggesting that a larger patch size introduces an inherent compression-related degradation that cannot be fully recovered by a better VAE configuration. 
In contrast, further reducing the actual compression ratio to 1024 (patch size = 1) substantially improves both metrics, yielding the best PER of 15.0\% and the highest Quality score of 3.2. 
Overall, these results suggest that the effective token compression ratio is the dominant factor in balancing efficiency and audio quality. 
However, we did not further reduce the compression ratio because a too-low compression rate would cause the number of tokens to increase dramatically, leading to higher resource consumption and slower generation speed.

\begin{table}[htbp]
\centering
\caption{Ablation study on token compression ratio.}
\begin{tabular}{c|c|c|c|c}
\hline
\multicolumn{5}{c}{\textbf{PT-90s model}} \\
\hline
VAE compression ratio & patch size & Actual compression ratio & PER(\%) $\downarrow$ & Quality (1-5) $\uparrow$\\
\hline
2048 & 1 & 2048 & 19.2 & 2.1 \\
\hline
1024 & 2 & 2048 & 20.6 & 2.4\\
\hline
1024 & 1 & 1024 & \textbf{15.0} & \textbf{3.2}\\
\hline
\end{tabular}\label{table_vae_comp}
\end{table}
\section{Related Work}
Song/Music generation aims to produce coherent vocals and BGM given lyrics and style prompts. And there are many types of works implementing this basic target. For example, Musiclm(\cite{agostinelli2023musiclm}) generate music by multi-stage sequence-to-sequence pipeline. SongGen(\cite{liu2025songgen}) uses discrete tokens to train an autoregressive (AR) model. Yue(\cite{yuan2025yue}) and LeVo(\cite{lei2026levo}) also employ the discrete tokens to train multi-stage AR models, which are sometimes complex. SongCreator(\cite{lei2024songcreator}) employs U-Net based diffusion model, but fundamentally it still requires discrete tokens and a multi-stage generation pipeline. MusicCot(\cite{lam2025analyzable}) tries to use chain-of-thought to enhance the reasoning ability of songs. Recently, some works start to employ diffusion strategy to enhance the audio quality. DiffRhythm2(\cite{jiang2025diffrhythm}) proposes semi-autoregressive architecture based on block flow matching. ACE-Step1.5(\cite{gong2026ace}) proposes LM planner based diffusion model. They all require more elaborate training pipeline/component design. To this end, this paper proposes a simple yet effective pure diffusion framework that directly outputs demixed vocal and BGM stems for high-fidelity song generation. No gimmicky tricks are needed.
\section{Conclusion}
This paper presents WanSong v1.0, a simple yet effective pure diffusion-based framework for long-form, commercial-grade music generation. Unlike autoregressive systems and cascaded AR+diffusion designs, WanSong adopts an end-to-end diffusion pipeline where audio is modeled as continuous tokens and generated in a single stage, enabling both strong fidelity and efficient inference through step distillation.

A key insight of our work is that vocal and BGM should not be jointly encoded as entangled targets. To address the resulting imbalance and the failure of CFG to simultaneously preserve vocal accuracy and BGM strength, we introduce a dual-stem token scheme with joint learning within each block. This design fundamentally reduces mutual interference, allowing WanSong to directly output demixed vocals and BGM stems, which greatly simplifies downstream post-production and editing workflows.

In addition, we further improve listening quality by aligning the generation process with human preferences using RLHF, targeting multiple dimensions including musicality, lyric accuracy, and prompt alignment. Extensive evaluations on our WanSong benchmark demonstrate that WanSong achieves strong performance across objective metrics, validating the effectiveness of the proposed diffusion framework and dual-stem modeling strategy.

Overall, WanSong v1.0 establishes a practical direction for diffusion-based music foundation models—delivering high fidelity, multilingual capability, long-duration generation, and editable stem outputs—without relying on complicated, gimmicky components.

%\section{Authors}
%
%\newpage

\bibliographystyle{assets/plainnat}
\bibliography{main}

@String(ICASSP = {ICASSP})

@online{suno_discover,
  author       = {{Suno}},
  title        = {},
  url          = {https://suno.com/discover},
  year         = {2026},
  urldate      = {2026-07-07}
}

@online{mureka,
  author       = {{Mureka}},
  title        = {},
  url          = {https://www.mureka.cn/},
  year         = {2026},
  urldate      = {2026-07-07}
}

@misc{labs2025flux1kontextflowmatching,
      title={FLUX.1 Kontext: Flow Matching for In-Context Image Generation and Editing in Latent Space},
      author={Black Forest Labs and Stephen Batifol and Andreas Blattmann and Frederic Boesel and Saksham Consul and Cyril Diagne and Tim Dockhorn and Jack English and Zion English and Patrick Esser and Sumith Kulal and Kyle Lacey and Yam Levi and Cheng Li and Dominik Lorenz and Jonas Müller and Dustin Podell and Robin Rombach and Harry Saini and Axel Sauer and Luke Smith},
      year={2025},
      eprint={2506.15742},
      archivePrefix={arXiv},
      primaryClass={cs.GR},
      url={https://arxiv.org/abs/2506.15742},
}

@inproceedings{esser2024scaling,
  title={Scaling rectified flow transformers for high-resolution image synthesis},
  author={Esser, Patrick and Kulal, Sumith and Blattmann, Andreas and Entezari, Rahim and M{\"u}ller, Jonas and Saini, Harry and Levi, Yam and Lorenz, Dominik and Sauer, Axel and Boesel, Frederic and others},
  booktitle={Forty-first international conference on machine learning},
  year={2024}
}

@article{wan2025,
      title={Wan: Open and Advanced Large-Scale Video Generative Models}, 
      author={Team Wan and Ang Wang and Baole Ai and Bin Wen and Chaojie Mao and Chen-Wei Xie and Di Chen and Feiwu Yu and Haiming Zhao and Jianxiao Yang and Jianyuan Zeng and Jiayu Wang and Jingfeng Zhang and Jingren Zhou and Jinkai Wang and Jixuan Chen and Kai Zhu and Kang Zhao and Keyu Yan and Lianghua Huang and Mengyang Feng and Ningyi Zhang and Pandeng Li and Pingyu Wu and Ruihang Chu and Ruili Feng and Shiwei Zhang and Siyang Sun and Tao Fang and Tianxing Wang and Tianyi Gui and Tingyu Weng and Tong Shen and Wei Lin and Wei Wang and Wei Wang and Wenmeng Zhou and Wente Wang and Wenting Shen and Wenyuan Yu and Xianzhong Shi and Xiaoming Huang and Xin Xu and Yan Kou and Yangyu Lv and Yifei Li and Yijing Liu and Yiming Wang and Yingya Zhang and Yitong Huang and Yong Li and You Wu and Yu Liu and Yulin Pan and Yun Zheng and Yuntao Hong and Yupeng Shi and Yutong Feng and Zeyinzi Jiang and Zhen Han and Zhi-Fan Wu and Ziyu Liu},
      journal = {arXiv preprint arXiv:2503.20314},
      year={2025}
}

@article{lipman2022flow,
  title={Flow matching for generative modeling},
  author={Lipman, Yaron and Chen, Ricky TQ and Ben-Hamu, Heli and Nickel, Maximilian and Le, Matt},
  journal={arXiv preprint arXiv:2210.02747},
  year={2022}
}

@inproceedings{wallace2024diffusion,
  title={Diffusion model alignment using direct preference optimization},
  author={Wallace, Bram and Dang, Meihua and Rafailov, Rafael and Zhou, Linqi and Lou, Aaron and Purushwalkam, Senthil and Ermon, Stefano and Xiong, Caiming and Joty, Shafiq and Naik, Nikhil},
  booktitle={Proceedings of the IEEE/CVF Conference on Computer Vision and Pattern Recognition},
  pages={8228--8238},
  year={2024}
}

@article{xu2023imagereward,
  title={Imagereward: Learning and evaluating human preferences for text-to-image generation},
  author={Xu, Jiazheng and Liu, Xiao and Wu, Yuchen and Tong, Yuxuan and Li, Qinkai and Ding, Ming and Tang, Jie and Dong, Yuxiao},
  journal={Advances in Neural Information Processing Systems},
  volume={36},
  pages={15903--15935},
  year={2023}
}

@article{yao2025songeval,
  title={Songeval: A benchmark dataset for song aesthetics evaluation},
  author={Yao, Jixun and Ma, Guobin and Xue, Huixin and Chen, Huakang and Hao, Chunbo and Jiang, Yuepeng and Liu, Haohe and Yuan, Ruibin and Xu, Jin and Xue, Wei and others},
  journal={arXiv preprint arXiv:2505.10793},
  year={2025}
}

@article{lei2026levo,
  title={Levo: High-quality song generation with multi-preference alignment},
  author={Lei, Shun and Xu, Yaoxun and Zhang, Huaicheng and Chen, Hangting and Zhang, Yixuan and Yang, Chenyu and Zhu, Haina and Wang, Shuai and Wu, Zhiyong and Yu, Dong and others},
  journal={Advances in Neural Information Processing Systems},
  volume={38},
  pages={102448--102479},
  year={2026}
}

@article{zhu2025muq,
  title={Muq: Self-supervised music representation learning with mel residual vector quantization},
  author={Zhu, Haina and Zhou, Yizhi and Chen, Hangting and Yu, Jianwei and Ma, Ziyang and Gu, Rongzhi and Luo, Yi and Tan, Wei and Chen, Xie},
  journal={IEEE Transactions on Audio, Speech and Language Processing},
  year={2025},
  publisher={IEEE}
}

@article{agostinelli2023musiclm,
  title={Musiclm: Generating music from text},
  author={Agostinelli, Andrea and Denk, Timo I and Borsos, Zal{\'a}n and Engel, Jesse and Verzetti, Mauro and Caillon, Antoine and Huang, Qingqing and Jansen, Aren and Roberts, Adam and Tagliasacchi, Marco and others},
  journal={arXiv preprint arXiv:2301.11325},
  year={2023}
}

@article{liu2025songgen,
  title={Songgen: A single stage auto-regressive transformer for text-to-song generation},
  author={Liu, Zihan and Ding, Shuangrui and Zhang, Zhixiong and Dong, Xiaoyi and Zhang, Pan and Zang, Yuhang and Cao, Yuhang and Lin, Dahua and Wang, Jiaqi},
  journal={arXiv preprint arXiv:2502.13128},
  year={2025}
}

@article{yuan2025yue,
  title={Yue: Scaling open foundation models for long-form music generation},
  author={Yuan, Ruibin and Lin, Hanfeng and Guo, Shuyue and Zhang, Ge and Pan, Jiahao and Zang, Yongyi and Liu, Haohe and Liang, Yiming and Ma, Wenye and Du, Xingjian and others},
  journal={arXiv preprint arXiv:2503.08638},
  year={2025}
}

@article{lei2024songcreator,
  title={Songcreator: Lyrics-based universal song generation},
  author={Lei, Shun and Zhou, Yixuan and Tang, Boshi and Lam, Max W and Liu, Feng and Liu, Hangyu and Wu, Jingcheng and Kang, Shiyin and Wu, Zhiyong and Meng, Helen},
  journal={Advances in Neural Information Processing Systems},
  volume={37},
  pages={80107--80140},
  year={2024}
}

@article{lam2025analyzable,
  title={Analyzable chain-of-musical-thought prompting for high-fidelity music generation},
  author={Lam, Max WY and Xing, Yijin and You, Weiya and Wu, Jingcheng and Yin, Zongyu and Jiang, Fuqiang and Liu, Hangyu and Liu, Feng and Li, Xingda and Lu, Wei-Tsung and others},
  journal={arXiv preprint arXiv:2503.19611},
  year={2025}
}

@article{jiang2025diffrhythm,
  title={DiffRhythm 2: Efficient and High Fidelity Song Generation via Block Flow Matching},
  author={Jiang, Yuepeng and Chen, Huakang and Ning, Ziqian and Yao, Jixun and Han, Zerui and Wu, Di and Meng, Meng and Luan, Jian and Fu, Zhonghua and Xie, Lei},
  journal={arXiv preprint arXiv:2510.22950},
  year={2025}
}

@article{gong2026ace,
  title={Ace-step 1.5: Pushing the boundaries of open-source music generation},
  author={Gong, Junmin and Song, Yulin and Zhao, Wenxiao and Wang, Sen and Xu, Shengyuan and Guo, Jing and Yang, Xuerui},
  journal={arXiv preprint arXiv:2602.00744},
  year={2026}
}

@inproceedings{stableaudio,
  title={Stable audio open},
  author={Evans, Zach and Parker, Julian D and Carr, CJ and Zukowski, Zack and Taylor, Josiah and Pons, Jordi},
  booktitle={ICASSP},
  pages={1--5},
  year={2025},
  organization={IEEE}
}

@article{seedtts,
  title={Seed-tts: A family of high-quality versatile speech generation models},
  author={Anastassiou, Philip and Chen, Jiawei and Chen, Jitong and Chen, Yuanzhe and Chen, Zhuo and Chen, Ziyi and Cong, Jian and Deng, Lelai and Ding, Chuang and Gao, Lu and others},
  journal={arXiv preprint arXiv:2406.02430},
  year={2024}
}

@article{tjandra2025aes,
    title={Meta Audiobox Aesthetics: Unified Automatic Quality Assessment for Speech, Music, and Sound},
    author={Andros Tjandra and Yi-Chiao Wu and Baishan Guo and John Hoffman and Brian Ellis and Apoorv Vyas and Bowen Shi and Sanyuan Chen and Matt Le and Nick Zacharov and Carleigh Wood and Ann Lee and Wei-Ning Hsu},
    year={2025},
    url={https://arxiv.org/abs/2502.05139}
}

@article{wu2026songbench,
  title={SongBench: A Fine-Grained Multi-Aspect Benchmark for Song Quality Assessment},
  author={Wu, Dapeng and Lei, Shun and Tan, Wei and Li, Guangzheng and Wang, Yunzhe and Zhang, Huaicheng and Zuo, Lishi and Wu, Zhiyong},
  journal={arXiv preprint arXiv:2604.25937},
  year={2026}
}

% \clearpage
% \newpage
% \beginappendix
% \input{section/6_supp}

\end{document}